\documentstyle[11pt,newpasp,twoside,epsfig]{article}
\begin{document}
\title{The Peculiar Cooling Flow Cluster RX\,J0820.9+0752}
 \author{C. M. Bayer-Kim, C. S. Crawford, S. W. Allen, A. C. Fabian}
\affil{University of Cambridge, Institute of Astronomy, Madingley
Road, Cambridge, CB3 0HA, UK}
\author{A. C. Edge}
\affil{Department of Physics, University of Durham, South Road, Durham
DH1 3LE}

\begin{abstract}
This poster is a summary of a paper on the subject (Bayer-Kim et
al. 2002) that has been submitted for publication in MNRAS.
We present X-ray, H$\alpha$, radio, and optical data, both imaging and
spectroscopy, of the cluster RX\,J0820.9+0752 and its peculiar central
cluster galaxy (CCG). We announce the discovery of several isolated
off-nuclear patches of blue light with strong line-emssion (total
L(H$\alpha)\sim 10^{42}\mbox{ergs}^{-1}$) and distinct ionizational
and kinematic properties, embedded in a region of H$\alpha$/X-ray
emission. We propose  and investigate a scenario in which a secondary
galaxy also featured in our observations has moved through a cooling
wake produced by the CCG, thereby producing the observed clumpy
morphology and helping to trigger star-formation within the clumps.
\end{abstract}

\section{RX~J0820.9+0752 At Different Wavelengths}
From the strongly peaked profile in our 9.4\thinspace ks {\sl Chandra}
observation (Fig. 1a), we derive that the hot
intracluster gas in the cluster RX\,J0820.9+0752 is cooling within a
radius of $r\approx 20$\thinspace kpc. The mass deposition rate of
a few tens of solar masses per year is consistent with the
new, reduced cooling flow scenario proposed by Voigt et al. (2002) as a
possible solution to the problem of the missing soft
X-ray luminosity in cooling flow clusters (Peterson et al. 2001). The
X-ray emission
appears extended to the NW and is coincident with a luminous H$\alpha$
nebula seen in an AAT image taken through a narrow-band
filter in the H$\alpha$ light (Fig. 1b). The cluster contains
only a weak radio source (flux density of $0.45\pm 0.13$\thinspace mJy
at 4.89\thinspace GHz as determined from a complete VLA snapshot
campaign; Edge et al. 2002, in prep.), but in an HST image its central
galaxy shows
highly peculiar morphology with two arcs of clumped emission to the NE
and a secondary elliptical to the SE, opposite the X-ray/H$\alpha$
feature (Fig. 1c). Several other isolated blobs
are also apparent scattered further out to the NW and E.

\begin{figure}
\centering
\epsfig{file=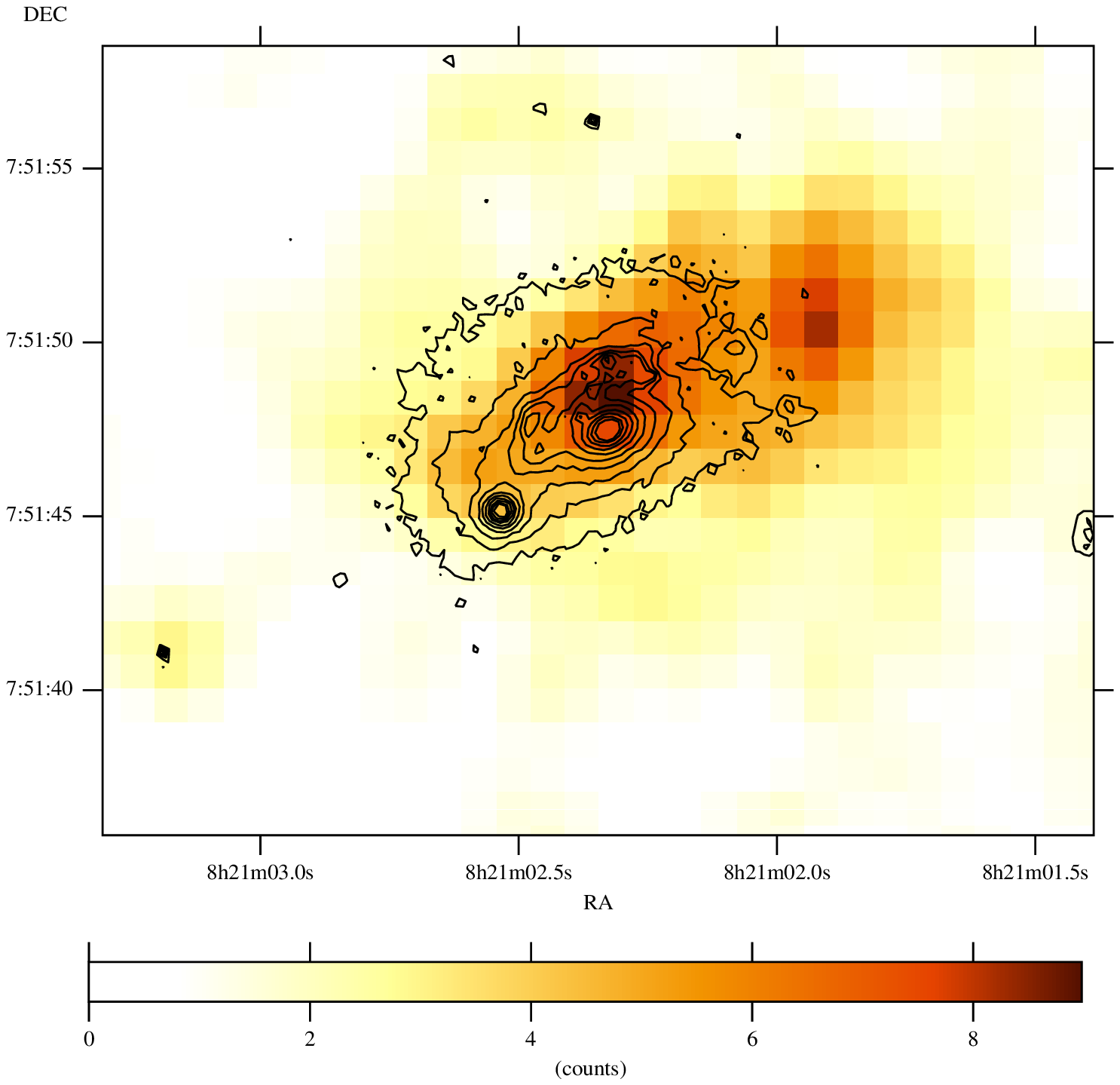,height=60truemm, width=90truemm}\\
\epsfig{file=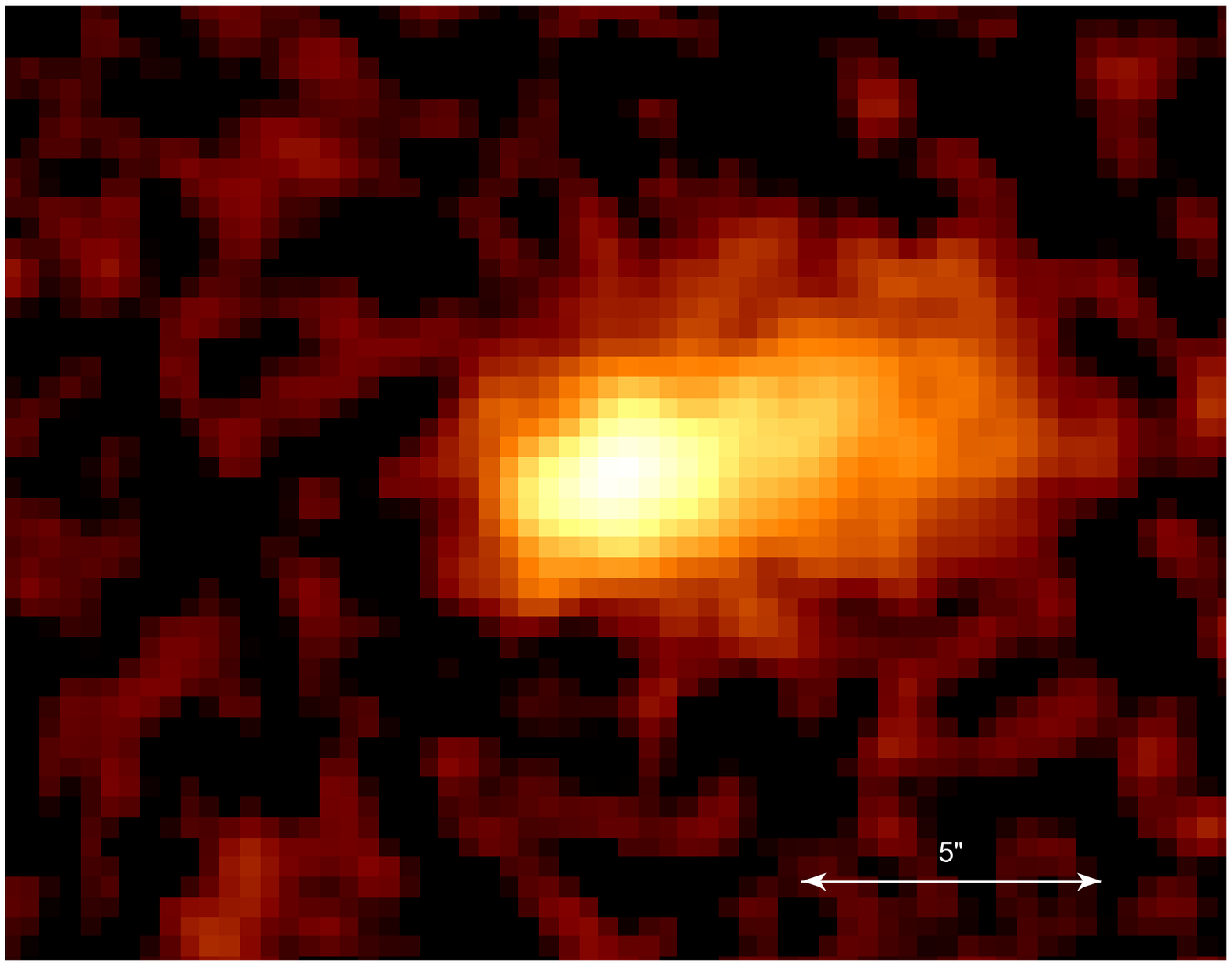,height=60truemm, width=90truemm}\\
\epsfig{file=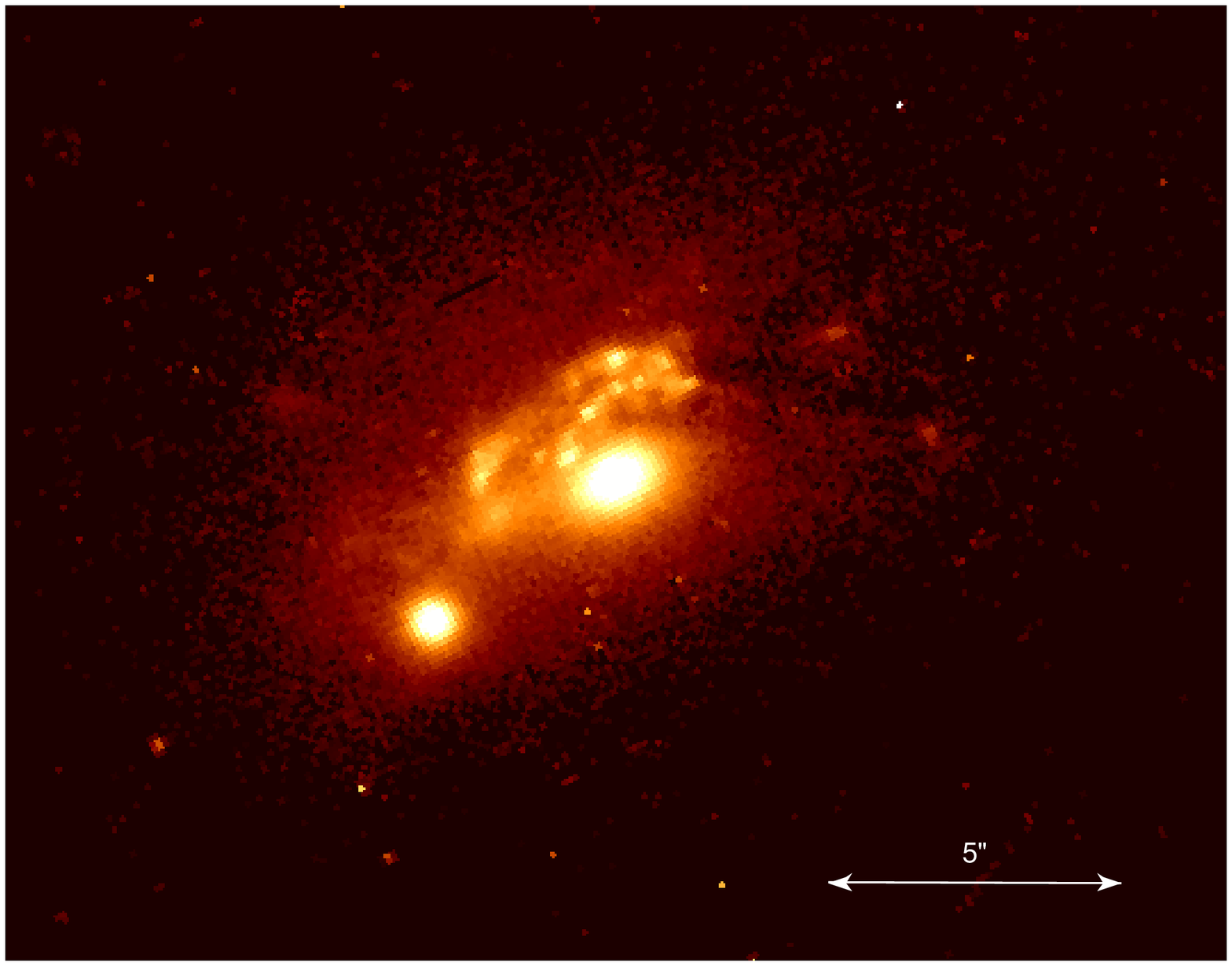,height=60truemm, width=90truemm}
\caption{(a) The slightly smoothed $0.5-2$\thinspace
keV {\sl Chandra} image of the cluster emission from RX~J0820.9+0752
with the optical contours overlaid (top). (b) Continuum-subtracted H$\alpha$
emission around the central
galaxy. The image shown is $20.5\times 16$ arcsec (middle). (c)
F606W HST image ($\Delta\lambda \approx 4490-5910$\AA\ at the redshift of the
object) of the central region of the cluster on the same scale
as the H$\alpha$ image (bottom). North is to the top and East to
the left in all the figures.}
\end{figure}


\section{Optical Spectroscopy}

We acquired optical spectra on the red ($3300-5000$\AA) and blue
($5100-6800$\AA{} rest wavelength) arms of ISIS (WHT, La Palma) to
cover both the known blue excess and the strong red line emission in
this object. The high spectral resolution allowed us to
extract separate spectra of some of the blobs seen in the HST
image. They are regions of very strong line emission (H$\alpha$ slit
Flux $\sim 10^{-15}$erg\thinspace s$^{-1}$\thinspace
cm$^{-2}$\thinspace\AA$^{-1}$), associated with a dust-rich
environment ($E(B-V)$ varies between about $0.3-0.7$ as determined
from the Balmer
decrement). Diagnostic line-ratios sensitive to ionization
properties show considerable variation throughout the whole object,
but with consistent deviations associated with the blobs
(Fig. 2a). The blobs are also kinematically distinct regions with
separate $v_{rad}$ and FWHM properties (Fig. 2b).

We find the blobs to be associated with very blue continua, whereas
the nuclei of the two galaxies are perfectly consistent with those of
``normal'' ellipticals. We have fit the blue blob spectra using models
constructed from an average template CCG (Crawford et al. 1999) and a variety
of MS stellar spectra from Pickles' (1998) stellar spectral flux
library (Fig. 3). It is possible to explain the blue excess by
the presence of
significant numbers of early stars (up to $\sim 10^5$ O-stars) from recent
star-bursts of varying ages.

\begin{figure}
\centering
\epsfig{file=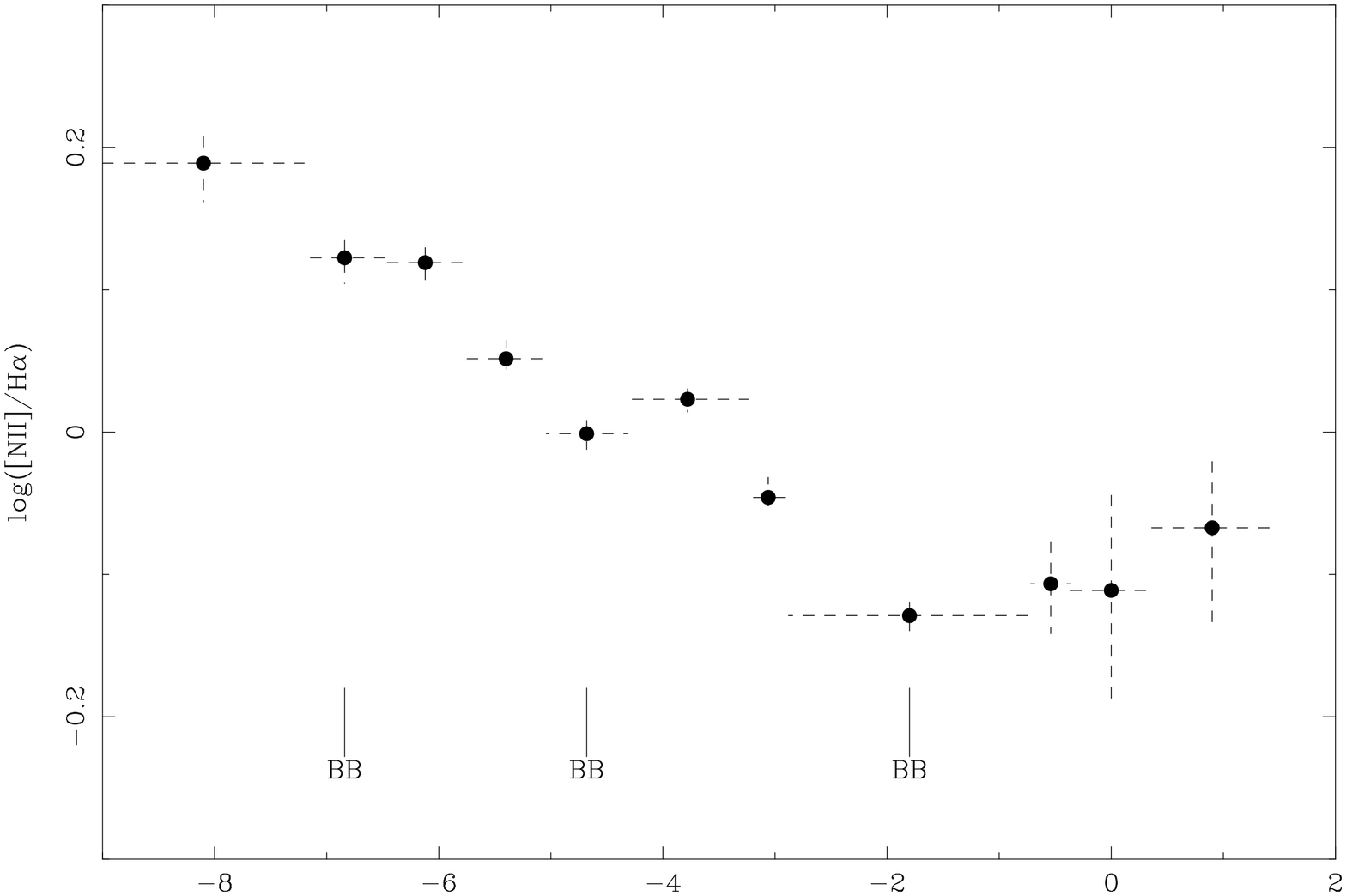,height=90truemm,width=45truemm,angle=270}\\
\epsfig{file=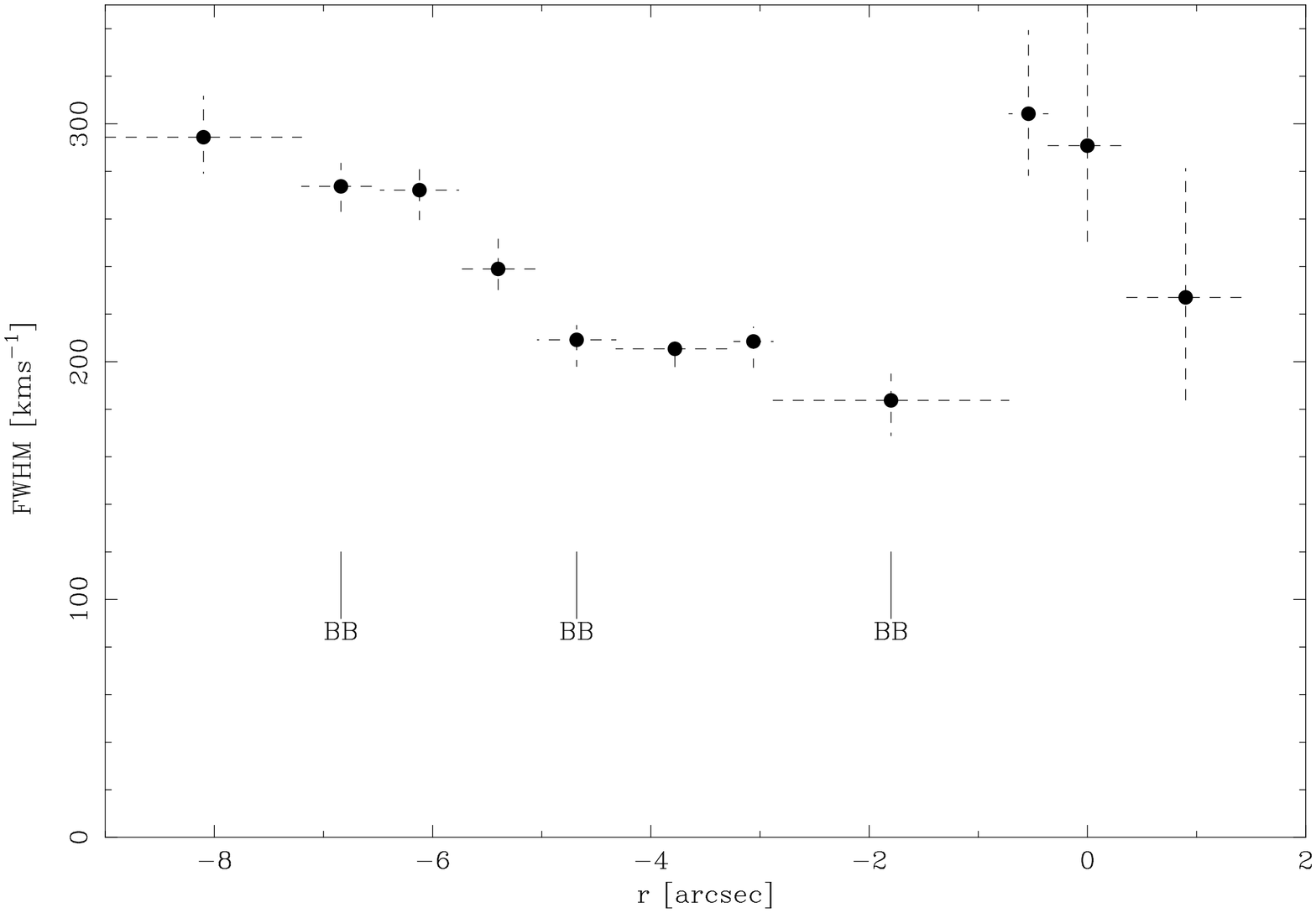,height=90truemm,width=45truemm,angle=270}
\caption{(a) Plot of [NII]/H$\alpha$ as a function of
radius. The positions of three blue blobs are marked. Note their
association with lower values of [NII]/H$\alpha$ (top). (b)
FWHM plotted vs r. The same blobs as above are denoted. Again they
show deviations from the general trend (bottom).}
\label{n2s2}
\end{figure}


\section{Conclusions}

The association of the H$\alpha$ nebula with the X-ray emission is
reminiscent of a similar feature found recently in A1795 (Fabian et
al. 2001b). The most
probable explanation given in this paper is that
the filament represents a ``cooling wake'', produced by the central
cluster galaxy moving through the hot ICM. We show that the same
explanation can be applied to RX\,J0820.9+0752, where the weakness of
the radio source practically rules out the possibility that it has
played a significant role in producing the wake. We find that the
energy output through the strong line-emission is consistent with
ionization from energetic photons produced by young stars. However, we
observe a velocity offset of up to 200\thinspace kms$^{-1}$ between
the young stellar components and the line-emitting gas in the
filament, suggesting they are not co-spatial. This could indicate that
massive stars are not the only ionization source for the nebula.
Maybe processes such as the ``cold mixing model'' proposed by Fabian
et al. (2001a) can account for at least part of the observed line-emission.

An important factor in producing the clumpy morphology and triggering the
star-bursts seems to be the secondary elliptical galaxy, whose
velocity properties relate it to the cooling gas. We propose a
scenario in which the secondary galaxy passes through the cooling wake
produced by the CCG.

\begin{figure}
\centering
\caption{Fit to one of the blue blobs. The spectrum is
represented by the dashed line, the model by the solid one. The dotted
spectrum is the difference
between the late component of the model and the blob spectrum, clearly
showing the blue excess and the Balmer and [OII]$\lambda 3727$\AA{}
emission lines.}
\epsfig{file=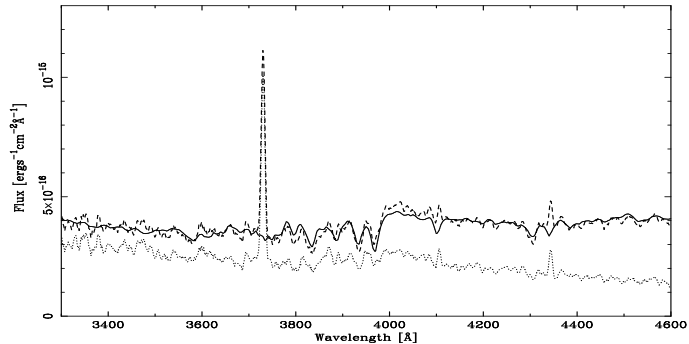,height=90truemm,width=45truemm,angle=270}
\label{Fit}
\end{figure}

\end{document}